\documentclass[	
						a4paper,
						showkeys,
					]{revtex4}

\usepackage[frenchb,english]{babel}
\usepackage[T1]{fontenc}
\usepackage{upgreek}
\usepackage{textcomp} 				
\usepackage[dvips]{graphicx}		
\usepackage{color}					
\usepackage[nice]{nicefrac} 		
\usepackage{amssymb,amsmath}		
\usepackage{array}					

\newcommand{\etal}{\textit{et al.}}

\begin{document}

\title{Pressure-temperature phase diagram of SrTiO$_3$ up to 53 GPa}
\date{\today}
\author{Mael Guennou}
\author{Pierre Bouvier}
\author{Jens Kreisel}
\affiliation{Laboratoire des Mat\'eriaux et du G\'enie Physique, CNRS - Grenoble Institute of Technology, MINATEC,
3 parvis Louis N\'eel, 38016 Grenoble, France}
\author{Denis Machon}
\affiliation{Laboratoire de Physique de la Mati\`ere Condens\'ee et Nanostructures, CNRS - Université Claude Bernard Lyon 1, 43 bvd du 11 Novembre 1918, 69622 Villeurbanne, France}

\begin{abstract}
We investigate the cubic to tetragonal phase transition in the pressure-temperature phase diagram of strontium titanate SrTiO$_3$ (STO) by means of Raman spectroscopy and X-ray diffraction on single crystal samples. X-ray diffraction experiments are performed at room temperature, 381 K and 467 K up to 53 GPa, 30 GPa and 26 GPa respectively. The observation of the superstructure reflections in the X-ray patterns provides evidence that the crystal undergoes at all investigated temperatures a pressure-induced transition from cubic to the tetragonal $I4/mcm$ phase, identical to the low-temperature phase. No other phase transition is observed at room temperature up to 53 GPa. Together with previously published data, our results allow us to propose a new linear phase boundary in the pressure-temperature phase diagram. The data are analyzed in the framework of the Landau theory of phase transitions. With a revised value of the coupling coefficient between the order parameter and the volume spontaneous strain, the model built from pressure-independent coefficients reproduces satisfactorily the boundary in the phase diagram, but fails at reflecting the more pronounced second-order character of the pressure-induced phase transition as compared to the temperature-induced transition. We propose a new Landau potential suitable for the description of the pressure-induced phase transition. Finally, we show that particular attention has to be paid to hydrostatic conditions in the study of the high-pressure phase transition in STO.
\end{abstract}


\maketitle

\section{Introduction}


The understanding of $AB$O$_3$ perovskite-type oxides is a very active research area with great relevance to both fundamental- and application-related issues. Perovskites form a particularly interesting class of materials because even slight modifications of its crystal structure can lead to drastic changes in physical properties, especially when the chemical substitution or external conditions, such as temperature or pressure, are altered. The ideal cubic structure of $AB$O$_3$ perovskites is essentially simple, with corner-linked anion octahedra $B$O$_6$, the $B$ cations at the center of the octahedra and the $A$ cations in the space between the octahedra. With respect to this ideal perovskite, structural distortions can be approximated by separating two main features \cite{Glazer1972,Glazer1975,Mitchell2002}: a rotation (tilt) of the $B$O$_6$ octahedra and/or cation displacements. Past investigations of these instabilities have been a rich source for the understanding of structural properties not only in perovskites but also in oxides in general. Namely, SrTiO$_3$ (STO) and BaTiO$_3$ have played an important role in the understanding of soft mode-driven phase transitions.

In the past, much progress in the understanding of perovskites has been achieved through temperature, electric field or chemical composition-dependent investigations. The use of the parameter pressure has been rarer due to inherent experimental difficulties, which have now been overcome for a number of years. Following the pioneering work by Samara \etal{} \cite{Samara1975}, it was generally accepted that pressure reduces zone-center ferroelectricity in $AB$O$_3$ perovskites, but increases anti-ferrodistortive (AFD) tilt instabilities at the zone-boundary. Until recently it seemed unlikely to discover any new fundamental properties and insight from pressure investigations of simple perovskites. However, within the last few years this perception has changed considerably, mainly following the discovery (i) that, in some perovskites, tilts can be suppressed under high-pressure \cite{Bouvier2002,Angel2005,Tohei2005}, (ii) that the effect of high pressure on the prototype ferroelectric PbTiO$_3$ is much more complex than was expected \cite{Kornev2005,Kornev2007,Janolin2008,Wu2005,Ahart2008} and (iii) that, according to ab-initio calculations \cite{Kornev2005,Kornev2007}, ferroelectricity is not suppressed by high-pressure in insulating perovskites but is found to be enhanced or induced as pressure increases above a critical value at very high pressure. Following this, the very recent literature has seen a renewal of high-pressure investigations into "simple" $AB$O$_3$ perovskites \cite{Kreisel2009}. There is nonetheless a great lack of comprehensive $P$-$T$ phase diagrams for perovskite and this although their complexity can provide a significant understanding of competing interactions. Notable exceptions concern the thorough investigation of the $P$-$T$ phase diagram of BaTiO$_3$ \cite{Ishidate1997,Zhong1994} or KNbO$_3$ \cite{Pruzan2007}.

Here we investigate strontium titanate SrTiO$_3$, which is generally considered to be a model perovskite and has been the object of constant attention for more than 40 years, both at very low temperatures for its quantum paraelectric behaviour and at higher temperatures for its ferroelastic AFD transition. At ambient conditions STO is cubic and its AFD transition to a tetragonal structure can be induced by decreasing temperature or increasing pressure. The temperature-induced transition has been intensely studied, both experimentally and theoretically, and is regarded as an archetype of a soft-mode-driven phase transition \cite{Fleury1968}. Several reviews \cite{Hayward1999,Carpenter2007,Cowley1996} have given comprehensive overviews on available experimental data in relation with the Landau theory of phase transitions. It was shown that the transition can been accurately described within the Landau framework by a 246 potential corrected for quantum saturation effects \cite{Carpenter2007}.

In contrast, the effect of hydrostatic pressure, the pressure-induced phase transition and thus the pressure-temperature phase diagram remain relatively unexplored, being limited to 15 GPa at room temperature. The pressure-induced phase transition in STO was first studied in the 70's by acoustic measurements at low temperatures \cite{Sorge1970,Fossheim1972,Okai1975} and the pressure-induced transition at room temperature was studied later by Brillouin \cite{Ishidate1988}, Raman spectroscopy \cite{Grzechnik1997,Lheureux2000} and acoustic measurements \cite{Lheureux2000a}. However, some questions in the Landau-type description remained unanswered. Especially, the pressure dependence of the coupling parameter between the order parameter and the spontaneous strain, directly related to the shape of the phase boundary in the $P$-$T$ phase diagram, is under question. Moreover, the possibility of a further phase transition to an orthorhombic phase at higher pressure has been raised \cite{Grzechnik1997} but not yet confirmed.

In this work, we carry out Raman scattering on STO single crystals at room temperature (RT). X-ray diffraction experiments are also performed at RT, 381 and 467 K. The aims of this study are (i) to clarify the $P$-$T$ phase diagram and the shape of the cubic-tetragonal phase boundary, (ii) to explore the phase diagram to higher pressures in view of potential phase transitions, and (iii) check the data against the proposed Landau potential describing the temperature-induced phase transition.

\section{Experimental details}

Experiments have been conducted on high-quality SrTiO$_3$ single crystals purchased from CrysTec GmbH. The samples were oriented along a $[110]_C$ direction (relative to the cubic unit cell) and polished to a thickness of about 10 $\upmu$m. All experiments were performed in diamond-anvil cells (DAC) with various pressure transmitting media, as detailed in the following. The cells have the Boehler-Almax diamond design with a cullet of 300 $\upmu$m.

X-ray diffraction experiments were performed on the ID27 beamline at the ESRF. The beam was monochromatic with a wavelength of 0.3738 \AA{} selected by an iodine K-edge filter and focused to a beam size of about 3 $\upmu$m. The signal was collected in the rotating crystal geometry (oscillation range of 64\degre{}) on a marCCD detector. For the room temperature measurement, neon was used as a pressure transmitting medium. The fluorescence of ruby was used as a pressure gauge \cite{Mao1978}. During our experiments, we have carefully followed the splitting between the two fluorescence lines R1 and R2 of ruby: the splitting retained a constant value of 1.42 nm up to 25 GPa, and then drifted slightly and reached 1.54 nm at 53 GPa, the highest pressure investigated, which indicates that the deviation from perfect hydrostatic conditions remains weak. In addition, we checked the pressure measurement by confronting the lattice constants of Neon to the curve calibrated by Dewaele \etal{} \cite{Dewaele2008}. Measurements were carried out with both increasing pressure and decreasing pressure, showing good reproducibility and no hysteresis. 

For measurements at high temperature, helium was used as pressure transmitting medium. The cell was heated with an external resistive heating device. A thermocouple was placed on the diamond for a rough adjustement of the temperature, and the final pressure and temperature were measured in the pressure chamber using a calibration method based on the fluoresence lines of ruby and SrB$_4$O$_7$:Sm$^{2+}$ \cite{Datchi1997}. A first measurement was carried out at 381(3) K (108\degres C) with increasing pressure from ambient pressure to 30 GPa and a second at 467(5) K (194\degres C) with decreasing pressure from 27 GPa to ambient pressure. At these temperatures, according to the melting curve determined by Datchi \etal{} \cite{Datchi2000}, helium remains liquid up to 17.6 GPa and 24 GPa respectively, providing perfectly hydrostatic conditions.

Raman scattering experiments were performed with Argon as a pressure transmitting medium. The spectra were recorded on a Labram spectrometer for the high-frequency part. For the low-frequency part, a separate measurement was performed on a T64000 Jobin-Yvon spectrometer with a cutoff frequency at 30 cm$^{-1}$. In both cases, the exciting laser line was the Argon at 514 nm with the incident laser propagating along a $[110]_C$ direction relative to the cubic axes. The laser power was kept at 10 mW on the DAC to avoid heating of the sample. 

\section{X-ray diffraction}


\begin{figure}[htbp]
\includegraphics[width=0.45\textwidth]{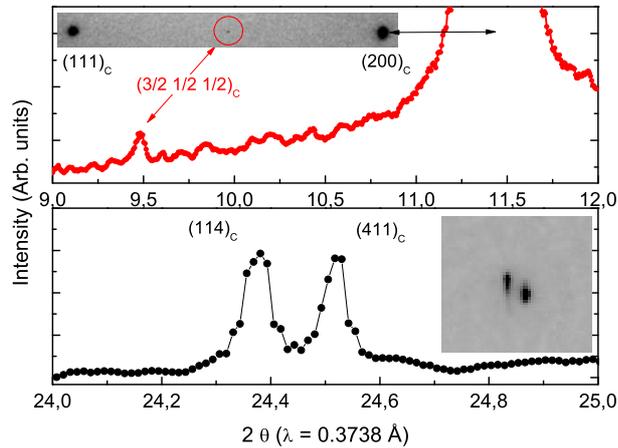}
\caption{Top: Example of a $(\frac{3}{2}\,\frac{1}{2}\,\frac{1}{2})_C$ superstructure reflection. Bottom: doubling of the $(411)_C$ reflection observed for a given orientation of the crystal, revealing the presence of the ferroelastic domain structure. Data are taken at 381 K and 30 GPa.}
\label{fig:detaildiffraction}
\end{figure}

Generally speaking, the changes of symmetry at the phase transition in perovskites can be followed by X-ray diffraction either by the splitting of Bragg peaks and/or by the emergence of superstructure reflections. In ferroic perovskites, distortions from the cubic unit cell are often very small, and the splitting of Bragg peaks hardly noticeable, so that the peak splitting is not always well suited for the phase determination unless high resolution diffraction setups are used. In our geometry, we could not detect a tetragonal distortion, defined as $|c/a-1|$, lower than 2.10$^{-3}$. In contrast, superstructure reflections are in our case a more efficient way to follow the transition. In perovskites where transitions are driven by the tilt of oxygen octaedra, such as in SrTiO$_3$, the intensities of the superstructure reflections are proportional to the tilt angle, so that their presence unambiguously proves the distortion from the cubic perovskite. In our experiments, we observed sets of superstructure peaks that are fully consistent with a transition from the $Pm\overline 3m$ to the $I4/mcm$ space group. The reflection conditions for the superstructure peaks indexed in the cubic unit cell $(\frac{h}{2}\,\frac{k}{2}\,\frac{l}{2})_C$ are $h, k, l = 2n+1$ and $h\neq k$ \cite{Glazer1975,Mitchell2002}. An example of an integrated curve (intensity vs. 2$\theta$) is given in figure \ref{fig:detaildiffraction} (top).

\begin{figure}[htbp]
\includegraphics[width=0.45\textwidth]{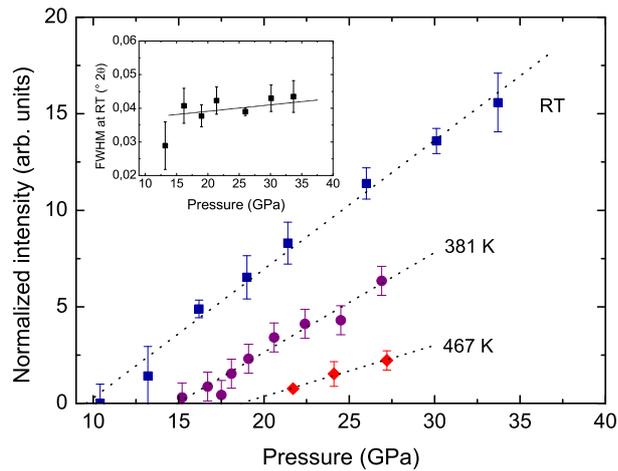}
\caption{Evolution of the $(\frac{5}{2}\,\frac{3}{2}\,\frac{1}{2})_C$ intensities as a function of pressure for the three temperatures investigated. The intensities are normalized by the intensity of the nearest Bragg peak from the (202) family and have been rescaled for clarity; they cannot be compared from one temperature to another. Insert: Evolution of the FWHM of the superstructure peaks at RT. The dotted line is a guide for the eye.}
\label{fig:surstructures}
\end{figure}

In the tetragonal phase, several ferroelastic domains can be expected in the crystal, with tetragonal $c$-axis directed along any of the three principal axes of the cubic phase. This domain structure complicates the analysis of the superstructure reflections since a reflection can be forbidden for a domain orientation and allowed in another. The observed intensity of the superstructure peaks for a given pressure then depends on the volume fraction probed for each domain, which may change from a measurement to another. During the measurements at high temperature, the domain structure of the crystal was clear from the doubling of many reflections; a typical example is shown in figure \ref{fig:detaildiffraction} (bottom). We avoided the ambiguity mentioned above by taking into account the reflections that are allowed for all domain orientations. For the $I4/mcm$ space group, those are the $(\frac{h}{2}\,\frac{k}{2}\,\frac{l}{2})_C$ with $h\neq k$, $k\neq l$ and $l\neq h$. The $(\frac{5}{2}\,\frac{3}{2}\,\frac{1}{2})_C$ is the first peak family that fulfills this condition. In the experiment at room temperature, the crystal did not show any sign of twinning upon decreasing pressure, after the crystal had been left in the cell for some time at high pressure, making this precaution in principle unnecessary. We follow nonetheless the same procedure in the following for the sake of consistency with the high-temperature measurements. The evolution with pressure of the normalized intensities at RT, 381 K and 467 K is shown in figure \ref{fig:surstructures}. In all cases, the intensities are normalized with respect to the intensity of the nearest $(202)_C$ Bragg peak. The evolution is linear within the pressure range investigated and therefore provides a reliable criterion for the determination of the transition pressure. We find $P_c =9.6$, 15.0 and 18.7 GPa at RT, 381 K and 467 K respectively. The full width at half maximum (FWHM) for the superstructure peaks at room temperature is shown in the insert ; it does not change appreciably within the pressure range investigated.

The lattice parameters were determined by a least square fit to the positions of the observed peaks within the cell aperture performed with the program UnitCell \cite{Holland1997}. For each pressure, between 20 and 80 peak positions were refined, including 10 to 15 superstructure reflections in the tetragonal phase. We retained for the fit only those peaks that could be indexed unambiguously in spite of the domain structure. The lattice constants obtained for measurements at room temperature, 381 K and 467 K are reported in table \ref{tab:latticeconstants}. Note that in a narrow pressure range above the transition (between 10 and 14 GPa at room temperature), the tetragonal distortion is too weak to be determined accurately although the presence of the superstructure reflections provides evidence that the transition to the tetragonal phase has occured.

\begin{table}[htbp]
\begin{center}
\begin{tabular}{l c m{3mm} l c c}
\hline\hline
\multicolumn{6}{c}{Room temperature}\\\hline
P (GPa) & $a_c$ & & P (GPa) & $a_{\mathrm{pc}}$ & $c_{\mathrm{pc}}$	 \\\cline{1-2}\cline{4-6}
0.72	& 3.9017(2)	& & 10.4	& 3.838(1) & 3.838(2) \\
1.27	& 3.8974(2)	& & 12.0	& 3.827(1) & 3.832(2) \\
2.15	& 3.8915(2)	& & 13.6	& 3.819(1) & 3.823(2) \\
2.73	& 3.8870(2)	& & 14.8	& 3.808(1) & 3.819(2) \\
3.21	& 3.8840(2)	& & 17.8	& 3.793(1) & 3.811(2) \\
3.68 	& 3.8796(2)	& & 19.5	& 3.784(1) & 3.803(2) \\
3.69 	& 3.8802(2)	& & 21.5	& 3.775(1) & 3.795(2) \\
4.39 	& 3.8756(2)	& & 23.9	& 3.761(1) & 3.789(2) \\
4.42 	& 3.8755(2)	& & 26.1	& 3.752(1) & 3.782(2) \\
4.80	& 3.8728(2)	& & 29.0	& 3.737(1) & 3.775(2) \\
5.64	& 3.8668(2)	& & 35.6	& 3.710(1) & 3.751(2) \\
5.80	& 3.8656(2)	& & 36.5	& 3.706(1) & 3.749(2) \\
6.01	& 3.8650(2)	& & 39.5	& 3.693(1) & 3.739(2) \\
6.42	& 3.8625(2)	& & 43.1	& 3.679(1) & 3.730(2) \\
6.93	& 3.8596(2)	& & 46.6	& 3.666(1) & 3.725(2) \\
7.08	& 3.8583(2)	& & 49.2	& 3.655(1) & 3.716(2) \\
7.46	& 3.8559(2)	& & 53.9	& 3.642(1) & 3.706(2) \\
7.85	& 3.8546(2)	& & 41.5	& 3.687(1) & 3.743(2) \\
8.17	& 3.8516(2)	& & 33.7	& 3.713(1) & 3.759(2) \\
8.67	& 3.8486(2)	& & 30.1	& 3.730(1) & 3.769(2) \\
8.88	& 3.8477(2)	& & 26.0	& 3.750(1) & 3.781(2) \\
9.40	& 3.8446(2)	& & 21.4	& 3.773(1) & 3.793(2) \\
		& 				& & 19.0	& 3.789(1) & 3.802(2) \\
		& 				& & 16.2	& 3.802(1) & 3.813(2) \\
		& 				& & 13.2	& 3.819(1) & 3.826(2) \\\hline
\multicolumn{6}{c}{381 K (108\degre C)}\\\hline
P (GPa) 	& $a_c$ & & P (GPa) & $a_{\mathrm{pc}}$ & $c_{\mathrm{pc}}$ \\\cline{1-2}\cline{4-6}
2.3	& 3.8969(5)	& & 16.7	& 3.809(3)$\phantom{0}$	& 3.809(3)$\phantom{0}$\\
5.7	& 3.8725(5)	& & 17.5	& 3.801(3)$\phantom{0}$	& 3.813(3)$\phantom{0}$\\
8.2	& 3.8552(5)	& & 18.1	& 3.795(3)$\phantom{0}$	& 3.814(3)$\phantom{0}$\\
9.8	& 3.8474(5)	& & 19.1	& 3.7943(9)	& 3.8015(9)	\\
11.8	& 3.8344(5)	& & 20.6	& 3.7861(9)	& 3.7964(9)	\\
12.9	& 3.8289(5)	& & 22.4	& 3.7759(9)	& 3.7894(9)	\\
14.2	& 3.8215(5)	& & 24.5	& 3.7642(9)	& 3.7809(9)	\\
15.2	& 3.8161(5)	& & 26.9	& 3.7521(9)	& 3.7723(9)	\\
		& 				& & 29.9	& 3.7374(9)	& 3.7631(9)	\\\hline
\multicolumn{6}{c}{467 K (194\degre C)}\\\hline
P (GPa) 	& $a_c$ & & P (GPa) & $a_{\mathrm{pc}}$ & $c_{\mathrm{pc}}$	 \\\cline{1-2}\cline{4-6}
2.5		& 3.9000(5)	& & 21.7 & 3.7852(9) & 3.7845(7) \\
9.8		& 3.8503(5)	& & 24.1 & 3.7697(9)	& 3.7773(7) \\
12.4		& 3.8356(5)	& & 27.2 & 3.7577(9)	& 3.7702(7) \\
15.3	& 3.8195(5)	\\
19.4	& 3.7974(5)	\\
\hline\hline
\end{tabular}
\caption{Lattice constants as a function of pressure at room temperature, 381 K and 467 K. In the tetragonal phase, the lattice constants are given as pseudo-cubic lattice constants, which are related to the tetragonal lattice constants $a_t$ and $c_t$ by $a_{\mathrm{pc}} = a_t/\sqrt{2}$ and $c_{\mathrm{pc}} = c_t/2$.}
\label{tab:latticeconstants}
\end{center}
\end{table}

\begin{figure}[htbp]
\includegraphics[width=0.45\textwidth]{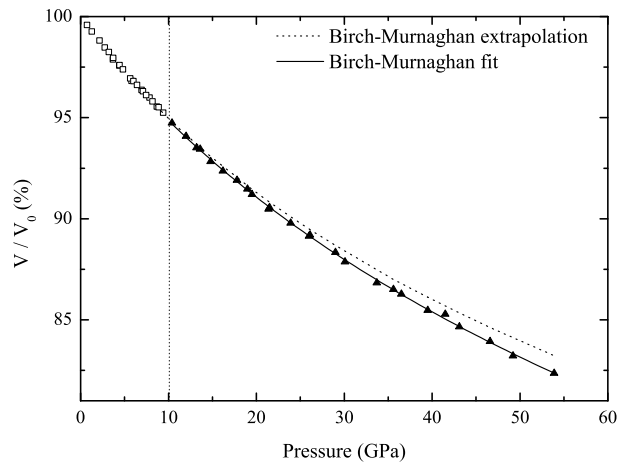}
\caption{Evolution of the relative volume with pressure at room temperature. The volume at ambient pressure ($V_0$=59.66(1) \AA$^{3}$) is taken from the fit to the Birch-Murnaghan EoS. The extrapolation calculated using the Birch-Murnaghan EoS is shown (dotted line), as well as the Birch-Murnaghan fit to the data in the tetragonal phase (solid line).}
\label{fig:volume}
\end{figure}
\begin{figure}[htbp]
\includegraphics[width=0.45\textwidth]{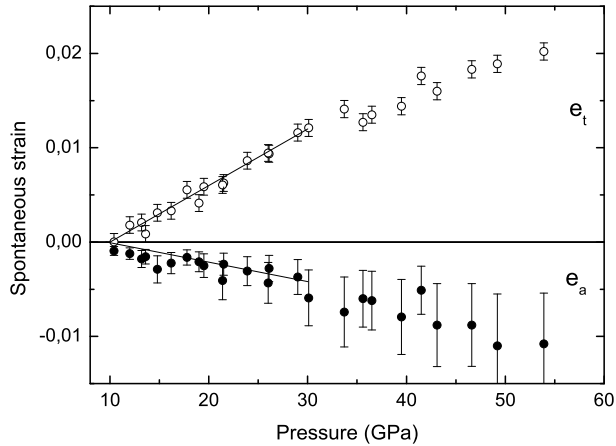}
\caption{Volume and tetragonal spontaneous strains $e_a$ and $e_t$ determined from the X-ray diffraction data at RT. The straight lines are least square fits to the data.}
\label{fig:spontaneousstrain}
\end{figure}

The evolution of the volume with pressure at room temperature is reported in figure \ref{fig:volume}. The plot of the generalized stress against the eulerian strain (the so-called $f$-$F$ plot \cite{Jeanloz1991}, not shown) is linear, showing that the description by a third-order Birch-Murnaghan equation of state (see e.g. \cite{Holzapfel1996}) in the cubic phase is appropriate. This fit yields values of $K_0=165(3)$ GPa and $K'_0=6.4(8)$ for the bulk modulus and its first derivative respectively, with $V_0=59.66(1)$. For comparison with previously published values \cite{Edwards1970,Ishidate1988,Beattie1971}, a fit with a Murnaghan equation of state (EoS) was also performed. The comparison is given in table \ref{tab:eos}. Our value for $K_0$ is slightly lower than the previously obtained values, which is accounted for by the more limited pressure range that was available in the previous studies: a fit of a Murnaghan EoS to our data in the 0-6 GPa range only yields $K_0=174(6)$ GPa. In addition, a Birch-Murnaghan fit was carried out in the tetragonal phase between 10 and 55 GPa, which gives a bulk modulus $K=225(6)$ GPa, derivative $K'=4.7(4)$ and a volume $V=56.59(4)$ \AA$^{3}$, all taken at 10 GPa. In order to calculate the spontaneous strains in the tetragonal phase, the volume of the cubic phase has to be extrapolated to the highest pressure measured using the EoS. The use of the simple Murnaghan EoS is prohibited given the high compressions at stake \cite{Holzapfel1996}. A way to limit the error on the extrapolated values is to use a EoS shifted to the transition pressure \cite{Holzapfel1996}. This fit, yields $K_{9.6} = 224(3)$ GPa and $K'_{9.6} = 5.9(6)$ with $V_{9.6} = 56.771(9)$ \AA$^{3}$. The extrapolation is shown in figure \ref{fig:volume}. We then make use of symmetry adapted spontaneous strains defined by $e_a = (2e_1 + e_3)$ (volume strain) and $e_t = 2(e_3-e_2)/\sqrt{3}$ (tetragonal strain) where $e_1$ and $e_3$ are the usual spontaneous strain components calculated from the lattice constant of the cubic phase $a_0$ extrapolated from the EoS and the pseudo-cubic lattice constants $a_{\mathrm{pc}}$ and $c_{\mathrm{pc}}$ by $e_1=e_2=(a_{\mathrm{pc}}-a_0)/a_0$ and $e_3=(c_{\mathrm{pc}}-a_0)/a_0$. Figure \ref{fig:spontaneousstrain} shows the evolution of the calculated spontaneous strains with pressure. 

\begin{table}[htbp]
\begin{center}
\begin{tabular}{m{1.5cm} m{2cm} m{1.5cm} l l}
\hline\hline
EoS & $K_0$ (GPa) & $K'_0$ & $P$ range (GPa) & Reference \\\hline
M & 176 	& 4.4 & 0 - 2 & Edwards \etal{} \cite{Edwards1970}\\
M & 172.1 & 5.81 & 0 - 2.2 & Beattie \etal{} \cite{Beattie1971}\\
M & 171.6 & 5.25 & 0 - 6 & Ishidate \etal{} \cite{Ishidate1988}\\
M & 166(3) & 6.2(7) & 0 - 9.5 & This work\\
BM & 165(3) & 6.4(8) & 0 - 9.5 & This work\\
\hline\hline
\end{tabular}
\caption{Bulk modulus $K_0$ and derivative $K'_0$ obtained from fits using a Murnaghan (M) or Birch-Murnaghan (BM) equations of state (EoS).}
\label{tab:eos}
\end{center}
\end{table}

\section{Raman scattering}

In the cubic phase of SrTiO$_3$, first-order Raman scattering is forbidden by symmetry. Nevertheless, a strong and broad scattering signature is observed, in agreement with the literature\cite{Perry1967, Nilsen1968,Ishidate1992,Grzechnik1997}. Although the underlying mechanism of this broad scattering is not yet fully understood, it is generally accepted that it can be assigned to second-order features \cite{Perry1967, Nilsen1968}. 

In contrast to the cubic phase, first-order Raman-active modes emerge in the tetragonal phase. The symmetry analysis of Raman-active modes of SrTiO$_3$ in the $I4/mcm$ space group has been done previously and can be found for example in \cite{Petzelt2001}: There are 7 Raman-active modes that decompose into $\Gamma = 1 (\mathrm A_{1g} + \mathrm E_g) + 2 (\mathrm B_{1g} +\mathrm E_g) + 1\mathrm B_{2g}$ where modes resulting from the splitting of a parent triply degenerate mode in the cubic phase are grouped in parentheses. The two $(\mathrm A_{1g} + \mathrm E_g)$ are the soft modes that drive the AFD phase transition.

\begin{figure}[htbp]
\begin{center}
\includegraphics[width=0.45\textwidth]{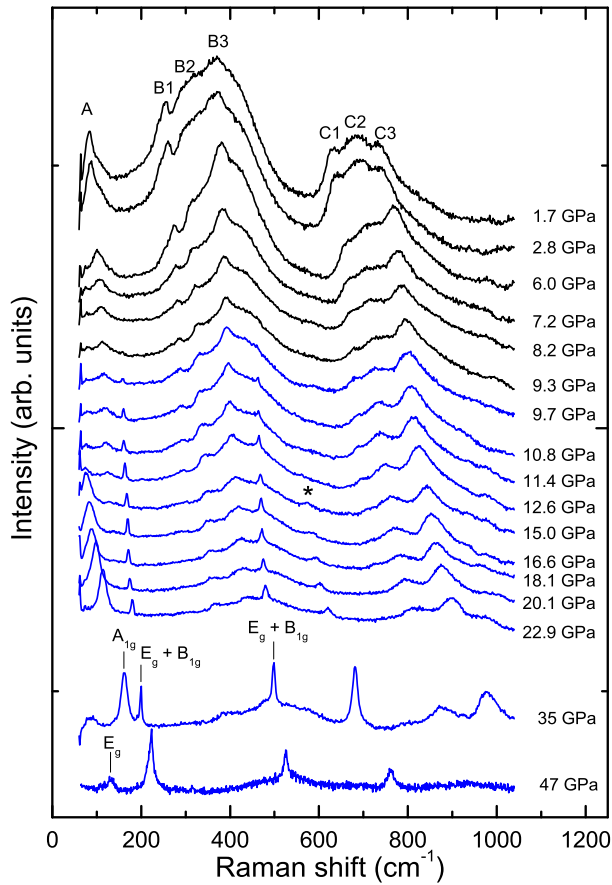}
\caption{Selection of Raman spectra recorded as a function of pressure at room temperature.}
\label{fig:RamanSpectra}
\end{center}
\end{figure}

\begin{figure}[htbp]
\begin{center}
\includegraphics[width=0.45\textwidth]{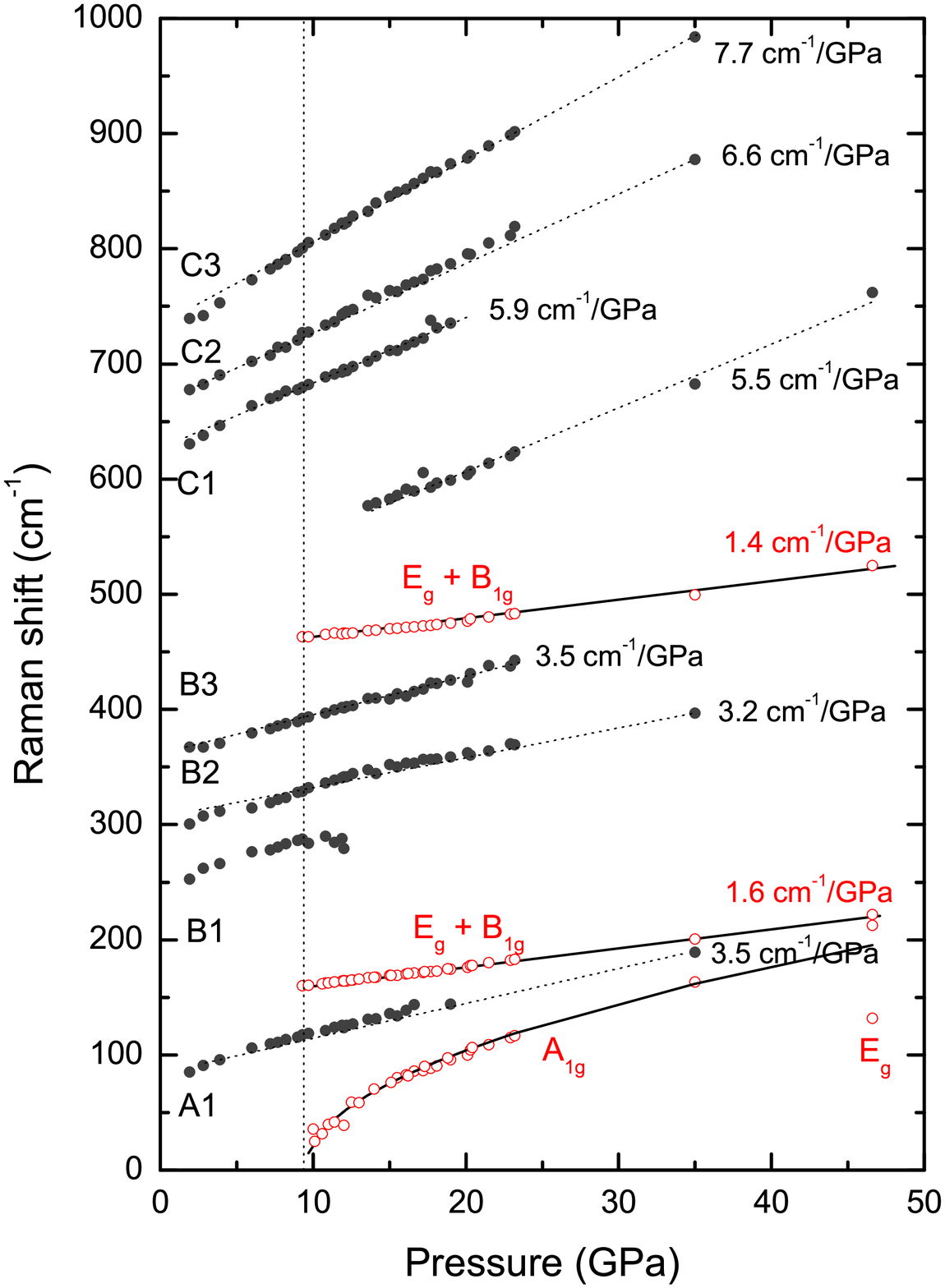}
\caption{Evolution of the Raman shifts with pressure for second-order bands (full symbols) and first-order modes from the tetragonal phase (open symbols).}
\label{fig:RamanShifts}
\end{center}
\end{figure}

A selection of representative spectra is presented in figure \ref{fig:RamanSpectra}, while the evolution in wavenumbers of the bands is reported in figure \ref{fig:RamanShifts}. Similarly to previous Raman scattering studies under pressure, we find that the second-order Raman spectrum at ambient pressure can be separated into distinct features. The dominant feature is the presence of two wide bumps. The first bump between 200 and 400 cm$^{-1}$ can be adequately described by three peaks labelled B1, B2 and B3 added to a wide background peak for which no physical significance is claimed. A second bump between 600 and 800 cm$^{-1}$ can be fitted with three distinct peaks labelled C1, C2 and C3. In addition, there is a sharper peak around 90 cm$^{-1}$ which we label A. These feature can be followed consistently as pressure increases and the band frequencies increase linearly with slopes ranging from 3.1 to 7.7 cm$^{-1}$/GPa as reported in figure \ref{fig:RamanShifts}. Remarkably, the Grüneisen parameter defined by $(K_0/\omega_{0})(\partial \omega / \partial P)$ is approximately identical for all second order features B and C ($\approx$ 1.6). 

As can be seen in figure \ref{fig:RamanSpectra}, the intensity of the broad features decreases significantly with increasing pressure, although some of the features are still observed at a much higher pressure than previously reported: some features are still strong up to 35 GPa, in constrast to earlier reports which claimed their disappearance at 15 GPa \cite{Grzechnik1997} and 13 GPa \cite{Ishidate1988}. We note however that the spectral signature of the C bands presents a visible change between 3 and 6 GPa. On the basis of our diffraction results, we believe that this change is not related to any change in the average crystal structure but might well be related to local changes as discussed by Itié \etal{} \cite{Itie2006}. 

The mode around 550 cm$^{-1}$ (marked $*$ in figure \ref{fig:RamanSpectra}) has to be considered separately due to its emergence as a relatively sharp peak at about 15 GPa without being associated with any other change in the Raman spectrum or the X-ray diffraction patterns. Its frequency increases with a slope that is characteristic of all the second-order features. We therefore hypothesize that this mode is a second order mode whose intensity increases from the emergence of a critical point under pressure in the two-phonon density of states \cite{Ho1997}.

The two first-order peaks with $(\mathrm B_{1g} +\mathrm E_g)$ symmetry emerge at 9.5 GPa revealing the transition to the tetragonal phase. Their frequencies are consistent with previous findings: a first peak at 160 and a second at 460 cm$^{-1}$. Their frequencies increase linearly with slopes of the order of 1.5 cm$^{-1}$/GPa, half of the minimal slopes measured for the second-order features. The increase remains linear up to the highest pressure measured (47 GPa). The soft mode of $\mathrm A_g$ symmetry  was followed down to 40 cm$^{-1}$ approximately at 10 GPa. A fit using the standard expression $\omega^2 = \omega_0^2 (P/P_c - 1)$ gives a transition pressure of $P_c=9.5$ GPa and a frequency $\omega_0=98.7$ cm$^{-1}$, consistent with the emergence of the $(\mathrm B_{1g} +\mathrm E_g)$ peaks at the very same pressure. The frequency of the soft mode of $\mathrm E_g$ symmetry was too low in frequency to be followed with increasing pressure, and could only be observed in the 47 GPa spectrum at 132 cm$^{-1}$. The Raman-active B$_{2g}$ mode was not observed, which is consistent with our geometry where the incident laser propagates along an $a$-axis of the tetragonal cell ($[110]_C$ with respect to the cubic axis). 

\section{Discussion}

\subsection{$P$-$T$ Phase diagram}

\begin{figure}[htbp]
\includegraphics[width=0.45\textwidth]{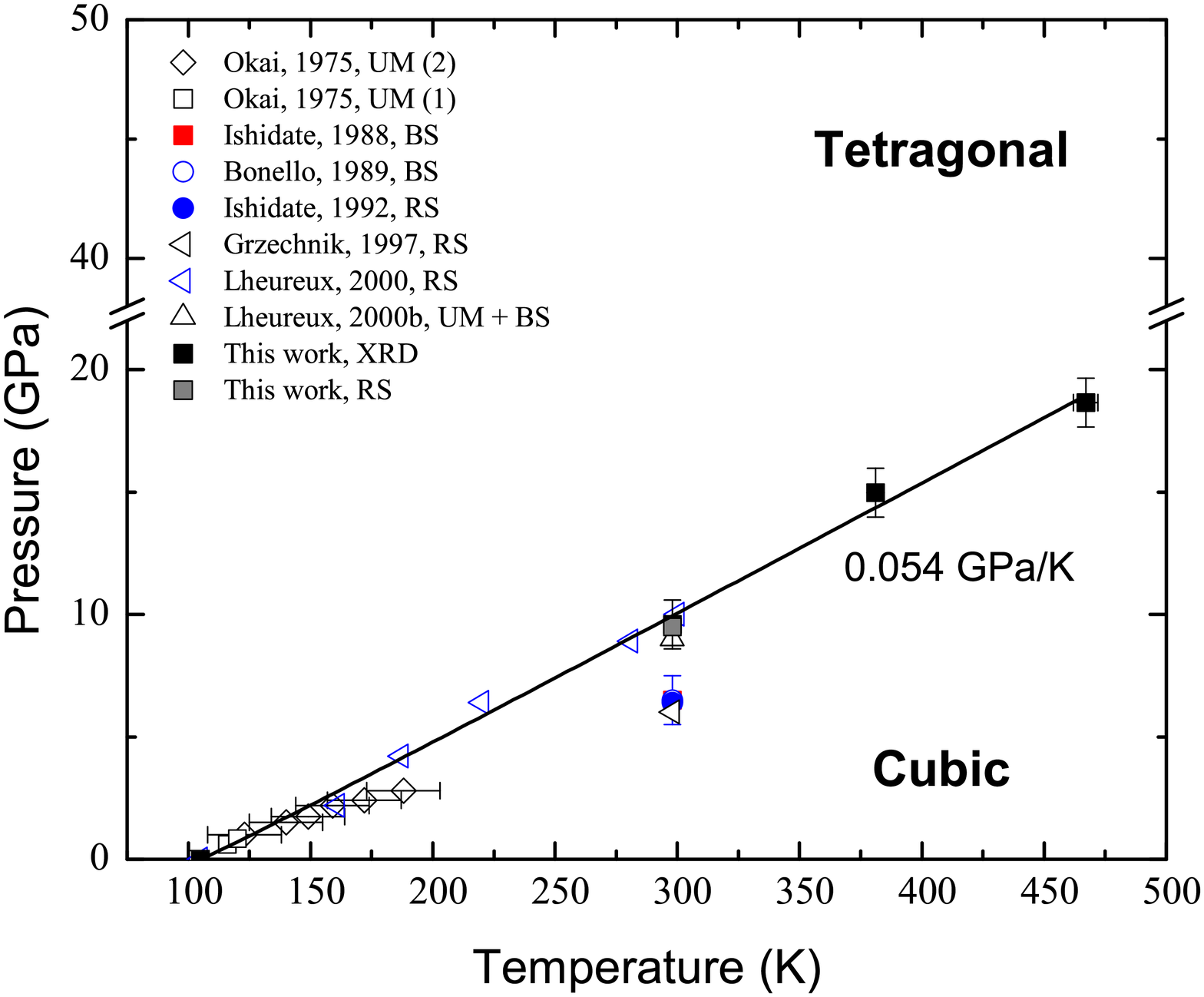}
\caption{Phase diagram for SrTiO$_3$ including our own results as well as previously published results obtained by ultrasonic measurements (UM), Brillouin scattering (BS) and Raman scattering (RS).}
\label{fig:phasediagram}
\end{figure}

Our data are reported in a $P$-$T$ phase diagram in figure \ref{fig:phasediagram} together with previous results from Okai and Yoshimoto \cite{Okai1975}, Grzechnik \etal{} \cite{Grzechnik1997} (Raman spectroscopy), Ishidate \etal{} \cite{Ishidate1992} (Brillouin spectroscopy) and Lheureux \etal{} \cite{Lheureux2000}. For the data by Okai \etal{}, we distinguish between low-pressure data (< 0.8 GPa) which they obtained with an Itskevich-type clamp technique and high-pressure data (> 0.8 GPa) obtained with a piston-cylinder device using the Teflon cell technique. 

The transition pressure at room temperature appears to be somewhat controversial. It was first determined by Ishidate \cite{Ishidate1988} \etal{} on the basis of Brillouin scattering measurements, where the authors estimated the transition pressure from the evolution of the elastic constants. Another Brillouin scattering experiment by Bonello \cite{Bonello1989} \etal{} was found consistent with this first estimation, as well as a later Raman scattering experiment by Ishidate \cite{Ishidate1992} \etal{}. A second Raman spectroscopy study was published by Grzechnik \cite{Grzechnik1997} \etal{} where they obtained the transition pressure by a fit of the soft mode frequency using a relation $\omega^2 = \omega_0^2 (P/P_c -1)$. Their fit yielded a critical pressure $P_c=6.04$ GPa, close to the value by Ishidate, although their values for $\omega_0$ differ significantly ($67.8$ cm$^{-1}$ vs. $82.4$ cm$^{-1}$ by Ishidate). A transition at 6 GPa was also reported by Lheureux \cite{Lheureux1999} \etal{} using a new ultrasonic measurement setup, but they revised their value to 9 GPa in a later study on the same crystals by ultrasonic measurements and Brillouin scattering \cite{Lheureux2000a}. Finally, high-pressure EXAFS \cite{Fischer1990} and XANES \cite{Itie2006} experiments suggest a local rearrangement of the Ti-cation displacement, which does not necessarily affect the average structure. 

In this work, we find a transition pressure of 9.5 GPa at room temperature from Raman scattering (emergence of first-order modes, evolution of the soft mode) and at 9.6 GPa from X-ray diffraction (evolution of the superstructure peaks). Our data show no evidence for a transition in the average structure at 6 GPa. Contrary to earlier work by Grzechnik \etal{}, we find at room temperature no indication for a further transition to an orthorhombic (or lower symmetry) phase at higher pressure up to 53 GPa.

Our data can be considered to a good approximation lying on a straight line passing through the zero-pressure transition at 105 K with a slope $\mathrm dP_c/\mathrm dT =$ 0.054 GPa/K. The Raman scattering (at low temperatures) and ultrasonics results (at room temperature) by Lheureux are in good agreement with this slope as well as the low-pressure measurements (below 1 GPa) by Okai \etal{}. Our results are however in disagreement with the higher pressure measurements by Okai \etal{}, whereby the authors measured the Brillouin shifts with decreasing temperature at a given pressure. A detailed re-examination of the data by Okai shows that the changes in the sound velocity are not abrupt but take place within a range of 10 to 20\degre C, leading to an according accuracy for a given pressure. We therefore do not regard this disagreement as significant. Another possible origin of the transition pressures reported by different authors will be commented on in the next section.

\subsection{Landau theory}

In the following, we will analyze our results within the framework of Landau theory, using the model recently summarized by Carpenter for STO \cite{Carpenter2007}, which satisfactorily accounts in a self consistent way for the temperature-induced phase transition. Our analysis will proceed in four steps. First, we will revise this model in the light of our new phase diagram and propose a new value for the coupling coefficient between the order parameter and the volume spontaneous strain. Second, the revised model will be tested against the high-pressure data; the limits of this model will be shown. Third, we will determine a new set of coefficients suitable for the description of the pressure-induced transition. Finally, we want to make use of the Landau approach to estimate the influence of anisotropic stress on the transition pressure.

\subsubsection{Determination of the coupling coefficient}

The potential summarized in \cite{Carpenter2007} consists of three terms: a 246 Landau development, a linear-quadratic coupling between the order parameter and the spontaneous strains and a pure elastic energy term. In the following, we restrict ourselves to the terms in the Landau potential that are relevant for the description of the $Pm\overline 3m$ to $I4/mcm$ phase transition and include the hydrostatic pressure. The order parameter $Q$ is associated to the softening of one component of a three-dimensional phonon mode at the R-point of the Brillouin zone. The quantum saturation effects at low temperatures are taken into account according to \cite{Salje1991} by introducing a saturation temperature $\Theta_s$ that characterizes the extent of the quantum mechanical regime. The general potential reads
\begin{eqnarray}
G & = & \frac{1}{2}\,A\,\Theta_{s} \left[\coth\frac{\Theta_{s}}{T} - \coth\frac{\Theta_{s}}{T_{c}}\right]Q^2 + \frac{1}{4}\,B\,Q^4 + \frac{1}{6}\,C\,Q^6 \nonumber\\
 & & + \lambda_2\,e_a\,Q^2 + 2\lambda_4\,e_t\,Q^2\nonumber\\
 & & + \frac{1}{4}(C_{11}^0-C_{12}^0)e_t^2 + \frac{1}{6}(C_{11}^0+2C_{12}^0)e_a^2\nonumber\\
 & & + P\,e_a \label{eq:landaupotential}
\end{eqnarray}
The values of all coefficients from \cite{Carpenter2007} are recalled in table \ref{tab:LandauCoefficients}. The $C_{\alpha\beta}^0$ are the elastic constants at ambient pressure in the cubic axes, $e_a$ and $e_t$ are the volume and tetragonal spontaneous strains respectively, $\lambda_2$ and $\lambda_4$ are the coupling coefficients between the order parameter and the spontaneous strains. Note that the combination $(C_{11}^0+2C_{12}^0)/3$ equals the bulk modulus $K$. The equilibrium values of the spontaneous strains are given by the condition $\partial G/\partial e = 0$. Substituting the spontaneous strains with their equilibrium values yields a renormalized 246 Landau potential $(A^*/2)\,Q^2 + (B^*/4)\,Q^4 + (C/6)\,Q^6$ with renormalized coefficients:
\begin{eqnarray}
A^* & = &  \,A\,\Theta_{s} \left[\coth\frac{\Theta_{s}}{T} - \coth\frac{\Theta_{s}}{T_{c}}\right] - \frac{2\lambda_2\,P}{\frac{1}{3}(C_{11}^0+2C_{12}^0)} \\
B^* & = &  B - \frac{2 \lambda_2^2}{\frac{1}{3}(C_{11}^0+2C_{12}^0)} - \frac{8\lambda_4^2}{\frac{1}{2}(C_{11}^0-C_{12}^0)} 
\end{eqnarray}

The values of the coefficients are determined from experimental observations. In particular, the best way to determine the coupling coefficient $\lambda_2$ is the examination of the phase boundary between the cubic and tetragonal phases in the $P$-$T$ phase diagram, defined by $A^*=0$. The expression of the boundary is 
\begin{equation}
P_c = A\,\Theta_s \frac{K}{2\lambda_2}\left[\coth\frac{\Theta_{s}}{T} - \coth\frac{\Theta_{s}}{T_{c}}\right]
\end{equation}
At temperatures higher than 200 K, this equation reduces to a linear phase boundary, with the assumption that $\lambda_2$ (or more generally the ratio $\lambda_2/K$) is pressure-independent. This point was questioned in \cite{Carpenter2007} in the light of the available data collected under pressure at that time. Our data however are fully consistent with this assumption. A least square fit of our transition pressures using this equation yields a ratio $\lambda_2/K$ of 1.67 10$^{-4}$, corresponding to $\lambda_2=0.030$ GPa with $K_0=180$ GPa, a markedly weaker coupling than the original value (0.046 GPa). This change has only minor consequences on the model. In particular, the renormalized coefficient $B^*$ remains almost unaltered by this change.

\subsubsection{Comparison against high-pressure data}

\begin{figure}[htbp]
\includegraphics[width=0.45\textwidth]{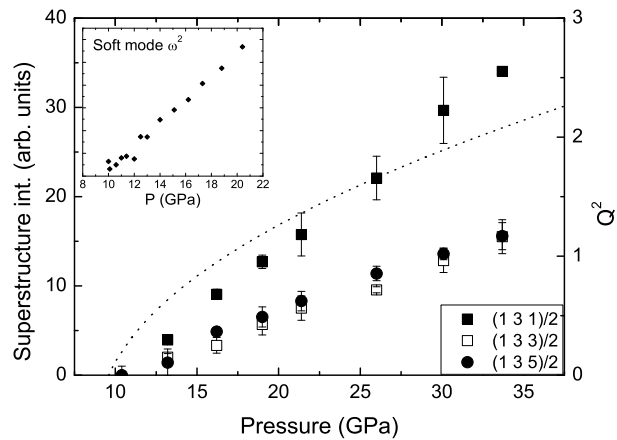}
\caption{Evolution of the intensities of different superstructure reflections. The order parameter squared $Q^2$ as calculated from equation (\ref{eq:orderparameter}) is plotted as the dotted line. The insert shows the evolution of the soft mode frequency squared $\omega^2$ measured by Raman spectroscopy.}
\label{fig:surstructure2}
\end{figure}

In the next step, we want to check this revised model against our data collected at the pressure-induced phase transition. A key point is the evolution of the order parameter with pressure. Within the Landau model, the explicit expression for the evolution of the order parameter above the transition pressure $P_c$ is given by \cite{Carpenter2007a}:
\begin{equation}
Q^2 = \frac{1}{2C}\left[-B^* + \sqrt{B^{*2} + 4 \frac{2\lambda_2}{K}\,C\,(P - P_c)}\right]
\label{eq:orderparameter}
\end{equation}
Experimentally, the intensity of the superstructure peaks is expected to scale directly with $Q^2$ \cite{Asbrink1996}. As mentioned earlier, the crystal at room temperature did not show any sign of twinning upon decreasing pressure, which enabled us to follow the superstructure intensities for the more intense $(\frac{1}{2}\,\frac{3}{2}\,\frac{1}{2})_C$ and $(\frac{1}{2}\,\frac{3}{2}\,\frac{3}{2})_C$ in addition to the $(\frac{1}{2}\,\frac{3}{2}\,\frac{5}{2})_C$ reported in figure \ref{fig:surstructures}. We show in figure \ref{fig:surstructure2} the evolution of their intensities vs. pressure up to 35 GPa and compare it to the evolution of $Q^2$ calculated from equation (\ref{eq:orderparameter}) with all coefficients except $\lambda_2$ being kept at the values given by Carpenter. It remains linear in this pressure range, and cannot be considered to scale with the pronounced parabolic evolution of $Q^2$ within more than a very narrow pressure range above the transition. In contrast, the intensity of superlattice reflections measured at low temperatures by Hünnefeld \cite{Hayward1999} was following $Q^2$ with reasonable agreement over the whole temperature range. 

The linear evolution of the order parameter with pressure, as well as the linear evolution of the soft mode frequency squared (see the insert in figure \ref{fig:surstructure2}), is usually the signature of a second-order transition. This would be a notable difference with the low-temperature transition that was shown to be very close to the critical point \cite{Salje1998,Hayward1999}. Within a 246 potential, this could be interpreted as a change in the importance of the elastic energy term that renormalizes the coefficient $B$. In STO, the study of the transition under an uniaxial stress applied along $[110]_C$ has shown that the transition got closer to second-order under the applied stress \cite{Gallardo1996}. The first to second-order crossover in PbTiO$_3$ under pressure \cite{Sani2002}, as well as the second-order character of the cubic-tetragonal transition in KNbO$_3$ under pressure \cite{Pruzan2007}, provides another example of a pressure-induced change in the nature of a phase transition. The comparison with STO is nonetheless not straightforward in so far as those two latter compounds exhibit zone-center (and not zone-boundary) instabilities. Moreover, it must be pointed out that while our data cover an extended pressure range above the transition, they do not focus very closely on the behavior in the immediate vicinity of the phase transition. 

It thus becomes clear that the pressure-induced behaviour of STO cannot be satisfactorily reproduced by the earlier proposed 246 model, which is not surprising in itself since it has been primarily developed for the description of a phase transition at ambient pressure. In the study of temperature-induced phase transitions, the bare elastic constants are only weakly temperature-dependent, whereas they are strongly pressure-dependent. Such a dependence is usually not taken into account in the Landau analysis of phase transitions in general, with the notable exception of the model by Tröster \etal{} \cite{Troester2002,Schranz2007}, which addresses these pressure-specific issues in the framework of the finite strain theory. Our data provide evidence that the pressure dependence cannot be overlooked for the study of the pressure-induced phase transition in STO.

\subsubsection{Determination of Landau coefficients for the pressure-induced phase transition}

The development of a unified Landau model suitable for the description of both the low-temperature and the pressure-induced phase transition is beyond the scope of this paper. However, we want to give a simplified description of the pressure-induced transition from our experimental data, starting from the known properties at RT and the transition pressure. Given the linear evolution with pressure of the superstructure intensities and the square of the soft mode frequency, we shall describe the transition as pure second-order, considered as a simplified and limiting case ($C=0$). In this hypothesis, the order parameter $Q^2$ scales with $P-P_c$, and the volume strain and tetragonal strain as a function of pressure read
\begin{eqnarray}
e_a & = & - 2 \left[\frac{\lambda_2}{\frac{1}{3}(C_{11}^0+2C_{12}^0)}\right]^2\frac{P-P_c}{B^*}\label{eq:easlope}\\
e_t & = & - 4\, \frac{\lambda_4}{\frac{1}{2}(C_{11}^0-C_{12}^0)}\frac{\lambda_2}{\frac{1}{3}(C_{11}^0+2C_{12}^0)}\frac{P-P_c}{B^*}
\label{eq:etslope}
\end{eqnarray} 
The elastic constants should be taken at the transition pressure (9.6 GPa). They can be estimated from the Brillouin scattering data under pressure of Ishidate \etal{} \cite{Ishidate1988}; we retain a bulk modulus of 230 GPa and a difference $(C_{11}-C_{12})/2$ of 135 GPa. The ratio $\lambda_2/K$ is fixed by the experimental linear phase boundary, which gives a new value of the coupling coefficient $\lambda_2$. We recalculate the coefficients  $B$ and $\lambda_4$ from the spontaneous strains. $\lambda_4$ is best calculated from equations (\ref{eq:easlope}) and (\ref{eq:etslope}): 
\begin{equation}
\left(\frac{\partial e_t}{\partial P}\right)\,\left(\frac{\partial e_a}{\partial P}\right)^{-1} = 2\, \frac{\lambda_4}{\lambda_2}\,\frac{\frac{1}{3}(C_{11}+2C_{12})}{\frac{1}{2}(C_{11}-C_{12})}.
\end{equation}
Equation (\ref{eq:etslope}) can then be used to calculate $B^*$, which in turn enables us to calculate $B$. The new set of coefficients is summarized in table \ref{tab:LandauCoefficients} and can be used for the description of the pressure-induced transition.

\begin{table}[htbp]
\begin{tabular}{>{$}l<{$} r@{ }l r@{ }l l}
\hline\hline
& \multicolumn{2}{c}{From \cite{Carpenter2007}} & \multicolumn{2}{c}{This work}\\
& \multicolumn{2}{c}{Ambient pressure} & \multicolumn{2}{c}{9.6 GPa}\\
\hline
A	 			& 1.8&10$^{-5}$ & 1.8&10$^{-5}$ & GPa.K$^{-1}$\\
\Theta_{s}	& 60.8 & & 60.8 & & K\\
T_{c} 		& 105.6 & & 105.6 & & K\\
B				& \hspace{3mm}1.2281 &10$^{-3}$  & 9.2 &10$^{-5}$ & GPa\\
B^*			& 8.1 &10$^{-4}$& 7.6 &10$^{-5}$ & GPa\\
C				& 1.092 &10$^{-3}$& - & & GPa\\
\lambda_2	& 4.6&10$^{-2}$ & \hspace{3mm}3.07&10$^{-2}$\hspace{3mm} & GPa\\
\lambda_4	& -7.5&10$^{-2}$ & -1.15&10$^{-2}$ & GPa\\
(C_{11}+2C_{12})/3		&  \multicolumn{1}{r}{180} & & 230 & & GPa\\
(C_{11}-C_{12})/2		&  \multicolumn{1}{r}{114} & & 135 & & GPa\\
\hline\hline
\end{tabular}
\caption{Coefficients of the Landau potential.}
\label{tab:LandauCoefficients}
\end{table}

\subsubsection{Effect of non-hydrostatic pressure conditions}

Last, we use the Landau model to estimate the influence of anisotropic stress on the transition pressure. As mentioned earlier, the transition pressures at room temperature reported in the literature differ significantly from one author to another, ranging from 6 to 10 GPa\cite{Grzechnik1997,Lheureux2000,Lheureux2000a,Ishidate1992,Bonello1989,Lheureux1999}. Anisotropic stress can be a reason for these scattered results. Other possible causes might involve point defects related to non-stoechiometry, or extended defects such as dislocations, which are known to play a significant role in perovskites in general. The particular relevance of the defects in STO crystals has been reflected in studies of strong surface effects, whereby the transition temperature at the surface can be shifted by not less that 45 K from the bulk value of 105 K \cite{Mishina2000,Salman2006}. We focus here on the influence of anistropic stress only, which can be estimated using the Landau approach. 

An additional uniaxial stress component $\sigma$, which we choose in this special case lying along the $c$-axis of the tetragonal phase, is taken into account in the theory by adding a term to the potential (\ref{eq:landaupotential}) for the corresponding energy. The energy term reads $-\sigma\,e_3$ which has to be transformed into symmetry compatible strains $-\sigma\,(e_a/3 + e_t/\sqrt{3})$ before substituting the strains with their equilibrium values. Solving $A^*=0$ then gives the transition pressure. The sensitivity of the system to this additional strain can be characterized by the derivative of the (hydrostatic) transition pressure $P_c$ with respect to the applied anisotropic stress:
\begin{equation}
\frac{\partial P_c}{\partial\sigma} = -\frac{1}{3}- \frac{2}{\sqrt{3}}\, \frac{\lambda_4}{\lambda_2}\,\frac{\frac{1}{3}(C_{11}^0+2C_{12}^0)}{\frac{1}{2}(C_{11}^0-C_{12}^0)}
\end{equation}
Within the second order approximation, the comparison with equations (\ref{eq:easlope}) and (\ref{eq:etslope}) shows that this expression reduces to
\begin{equation}
\frac{\partial P_c}{\partial\sigma} = -\frac{1}{3}+ \frac{1}{\sqrt{3}}\, \left(\frac{\partial e_t}{\partial P}\right)\,\left(\frac{\partial e_a}{\partial P}\right)^{-1}
\end{equation}
It is therefore possible to estimate the sensitivity of the system to anisotropic stress on the basis of the experimental observation of the spontaneous strains only, without a detailed knowledge of the coefficients of the potential. Our observations lead to a value of -1.9, suggesting that the system is sensitive to anisotropic stress, which in turns emphasizes that great attention should be paid to the hydrostaticity in the high-pressure measurements and their analysis. 

\section{Conclusion}

In this work, we have performed X-ray diffraction experiments at room temperature, 381 and 467 K up to 53 GPa, 30 GPa and 26 GPa respectively. The observation of the superstructure reflections in the X-ray patterns provides evidence that the crystal undergoes at all investigated temperatures a pressure-induced transition from cubic to the tetragonal $I4/mcm$ phase, identical to the low-temperature phase. The intensity of the superstructure reflections, directly related to the rotation of the oxygen octahedra, evolves linearly, which enables a reliable and precise determination of the transition pressure to 9.6, 15.0 and 18.7 GPa for the three temperatures investigated. In addition, Raman scattering experiments up to 47 GPa at room temperature has enabled us to follow the evolution of the soft mode frequency with pressure, as well as the first order Raman modes allowed in the tetragonal phase. The observed transition pressure is in agreement with the X-ray diffraction results. Together with previously published data, our results allow us to propose a new picture of the phase boundary in the $P$-$T$ phase diagram. Furthermore, we find no indication at room temperature for a further transition to an orthorhombic (or lower symmetry) phase at high pressures up to 54 GPa.

The data have been analyzed in the framework of the Landau theory of phase transitions. The analysis has shown that the model recently summarized by Carpenter satisfactorily accounts for the transition line in the phase diagram with a revised value of the coupling coefficient $\lambda_2$. This is remarkable since it does not take into account any pressure dependence of the elastic constants. The pressure-induced transition however, shows a strong second-order character that cannot be captured by this model that was designed for the almost tricritical transition at low temperature, unless an effect of pressure on the coefficients of the model is taken into account. Within the hypothesis of a pure second-order phase transition, we have proposed a new set of coefficients suitable for the description of all our experimental observations accompanying the pressure-induced phase transition. The experimental values provided here can also be used as a test for a future unified model. Finally, we have estimated the sensitivity of the transition pressure to an anisotropic stress and shown that particular attention has to be paid to hydrostatic conditions.

\section{Aknowlegements}

The authors are grateful for precious help from the ESRF staff, especially G. Garbarino and M. Mezouar at the ID27 Beamline as well as W. Crichton, J. Jacobs and M. Hanfland for the preparation of the diamond anvil cells. The authors also wish to thank J.-P. Itié (Soleil - Orsay) for fruitful discussions as well as E.K.H. Salje for his most illuminating comments on this work. Support from the French National Research Agency (ANR Blanc PROPER) is greatly acknowledged.

\bibliographystyle{aip}
\bibliography{biblio}

\end{document}